\documentclass[journal]{IEEEtran}
\usepackage{graphicx}
\usepackage{color}
\usepackage{placeins}
\usepackage{float}
\usepackage{hyperref}
\usepackage{tabularx,colortbl}
\usepackage{subcaption}
\usepackage{amsmath}
\usepackage{comment}
\begin{document}

\title{Popup Arrays for Large Space-Borne Apertures}

\author{Oren S. Mizrahi, Austin Fikes, Alan Truong, Fabian Wiesemüller, Sergio Pellegrino, and Ali Hajimiri
\thanks{ O.S.M., A.F., and A.H. are with the Department
of Electrical Engineering, California Institute of Technology, Pasadena,
CA, 91125 U.S.A.

A.T and S.P. are with the Graduate Aerospace Laboratories (GALCIT), California Institute of Technology, Pasadena, CA 91125, U.S.A.

F.W. is with the Aerial Robotics Laboratory, Imperial College London, London, SW7 2AZ, U.K.}}

\markboth{}{}

\maketitle

\begin{abstract}
Large apertures in space are critical for high-power and high-bandwidth applications spanning wireless power transfer (WPT) and communication, however progress on this front is stunted by the geometric limitations of rocket flight. Here, we present a light and flexible 10GHz array, which is composed of dipole antennas co-cured to a glass-fiber composite. The arrays are designed to dynamically conform to new shapes and to be flexible enough to fold completely flat, coil, and pop back up upon deployment. The design was chosen to be amenable to scalable, automated manufacturing - a requirement for the massive production necessary for large apertures. Moreover, the arrays passed the standard gamut of required space-qualification testing: the antennas can survive mechanical stress, extreme temperatures, high-frequency temperature cycling, and prolonged stowage in the flattened configuration. The elements exhibit excellent electromagnetic performance - with a return ratio better than -10dB over a bandwidth of 1.5GHz and a single lobe half-power beam width
 of greater than $110^{\circ}$ suitable for broad range beamforming and with excellent manufacturing consistency. Moreover, its mechanical durability vis-a-vis extreme temperatures and protracted stowage lends itself to demanding space applications. This lightweight and scalable array is equipped to serve a host of new space-based radio-frequency technologies and applications which leverage large, stowable and durable array apertures.
\end{abstract}

\section{Introduction}
Space is an increasingly ripe realm for RF applications including space solar power \cite{HashemiNature, fikes_SSPD1, ayling_MPO, abiri_SSPP, fikes_SSPP, gal_sheets, kelzenberg_ultralight}, communications \cite{kodheli2020satellite, desanctis_satellite, huang2020_satellite}, and the traditional remote sensing applications \cite{stephens2007remote, tapete2017trends}. These applications demand high bandwidth and/or high power efficiency but are currently limited by the aperture size that can be deployed \cite{saito_SAR}; large apertures in space are challenging because of the crucial requirement that they fit within the fairing of the launch vehicle. Previous attempts to innovate around this problem have resulted in foldable \cite{ISARA,lee_packaging}, inflatable \cite{Inflatable}, rollable \cite{HashemiNature}, and other \cite{santiago2013advances,fikes_JOM} aperture designs. However, these apertures are still limited to a maximum of a few square meters \cite{qi_large_ring, warren_deployable,oegerle_atlast,wang2020space}. This is insufficient for future high-bandwidth and high-power applications, such as high-throughput internet routing \cite{burleigh2019connectivity, wei_satellite}. Moreover, a large aperture sufficiently greater than the state-of-the-art is a fundamental requirement for meaningful space solar power, which proposes the collection of solar power in space, subsequent down-conversion to RF, and wireless power transfer to Earth by electronic focusing and beam-steering \cite{fikes_SSPD1,fikes_SSPP}. Small apertures simply cannot collect or focus the necessary power to make such a scheme feasible - large, flexible apertures are required.

Large, flexible apertures and the applications they enable are currently seeing much interest. Attention is devoted to their physical design \cite{HashemiNature}, beam-steering in non-planar scenarios \cite{MIZRAHI_FLEXIBLE_RECONSTRUCTION, FIKES_FRAMEWORK_MTT, HAJIMIRI_DYNAMIC_FOCUSING}, and economic efficiency at large scale \cite{mizrahi_TEA}. Indeed, critical to flexible array design is the existence of compatibly flexible radiator: these have also seen attention \cite{chiao2007flex_patch, hester2016inkjet_ant, gal_sheets,HashemiNature,APSURSI_FIKES}.

Previous attempts \cite{Kalra_2017,HashemiNature} to design flexible antennas compatible with space missions have not resulted in candidates that are simultaneously scalable and space-ready. Scalability is critical because it enables the leveraging of existing industrial and manufacturing infrastructure to produce the thousands or millions of antennas necessary to populate a large aperture without sacrificing absolute performance or consistency thereof. It also allows for production to take place at efficient costs. The requirement for space-readiness is clear: antennas need to survive extreme temperatures, high-frequency thermal cycling - sometimes several cycles a day for decades - and need to operate even after months being stowed in the rocket.

Such structures must be compatible with rolling and foiling, adding further constraints on their mechanical design. For instance, while a multi-layer design, where the electronics and antennas are built on two separately contiguous layers \cite{HashemiNature}, may be compatible with deployment, folding and rolling can impose shear stress, leading to separation of the two layers or other damage to the structure.

The challenge lies in that an antenna is perhaps the most sensitive component of a microwave system with respect to geometry. In this work, we present an antenna which balances the trade-off between microwave performance and space-ready mechanical design, based on \cite{APSURSI_FIKES}. This paper focuses on the space-readiness of this approach and expands on performance and manufacturing process, and discusses the results of various tests performed to demonstrate space-readiness.

\section{Design}
\subsection{Mechanical Design}
\begin{figure}[H]
\centering
    \includegraphics[width=0.49\textwidth]{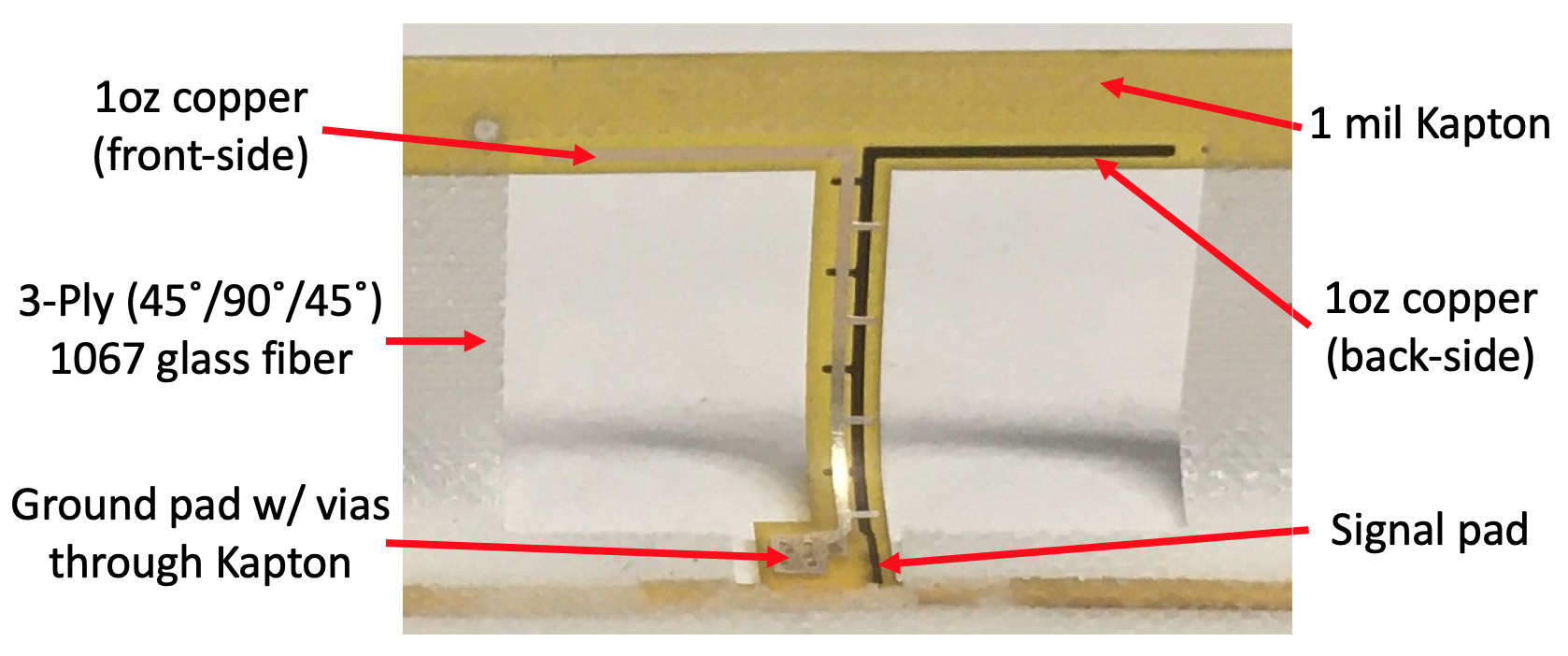}
    \caption{A collapsible dipole antenna in its operational configuration with labels.}
    \label{fig:antenna}
\end{figure}

\begin{figure}[H]
\centering
    \includegraphics[width=0.49\textwidth]{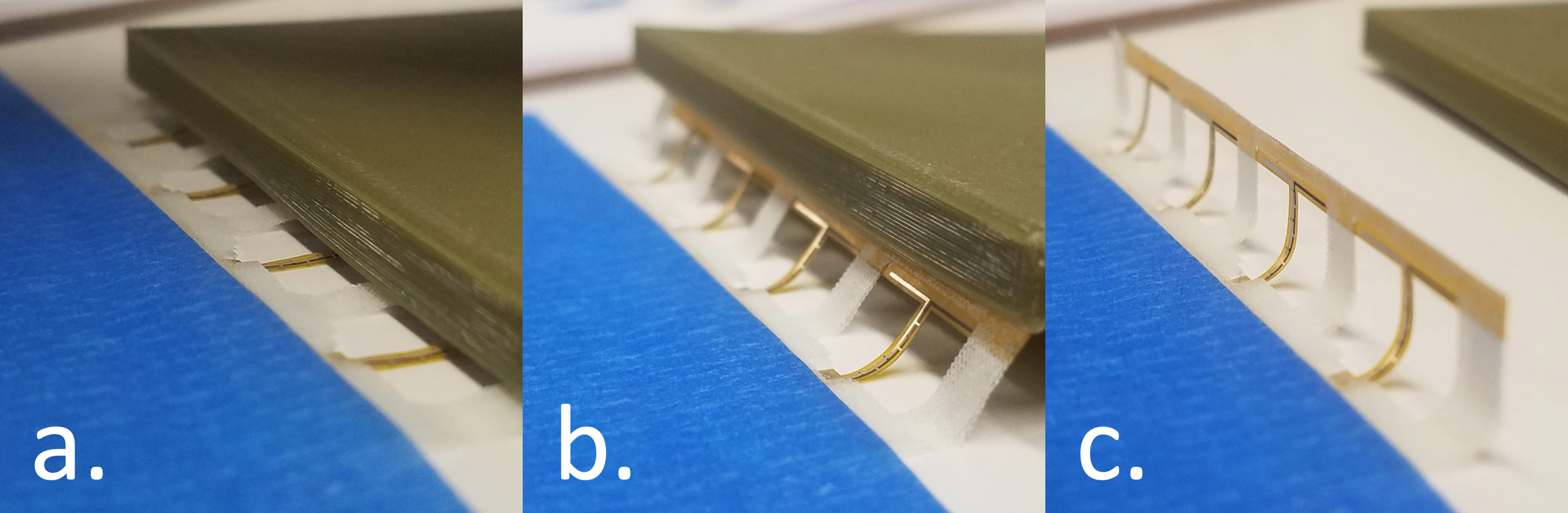}
    \caption{Antenna collapsibility: a. Collapsed, flat configuration, b. Intermediate state during redeployment. c. Operational configuration}
    \label{fig:flattening}
\end{figure}

The deployable dipole antenna is shown in Fig. \ref{fig:antenna}. Collapsibility and compatiblity with large-scale, lightweight arrays drive the shape, materials selection, and manufacturing process. The conductive portion of the antenna is made from a single 25$\mu$m-sheet of polyimide with etched 0.5oz copper on both sides. Use of thin polyimide, rather than traditional {\it rigid} substrates, facilitates the level of flexibility needed for collapsibility and self-deployment. The conductive antenna sheet is paired with a glass fiber composite frame, cured into the ``J" shape seen in Fig. \ref{fig:flattening}a, which provides the elastic deployment mechanism that pops up after flat stowage and also contributes the structural integrity necessary for consistent antenna shapes. The composite is a 3-ply stack of JPS Composites 1067 glass fiber impregnated with Patz-F4 resin which is layered in a $45^\circ / 90^\circ / 45^\circ$ configuration.

Various methods to adhere the conductive sheet to the glass fiber frame, such as use of epoxy, double-sided adhesive, or spray adhesive, do not produce reliable and repeatable results. They introduce unnecessary manufacturing variation, are difficult to scale, and negatively impact electromagnetic and mechanical performance. To circumvent these pitfalls, we used a technique called ``co-curing," which is widely used in the manufacturing of composite laminates but was not previously used for antenna design. Co-curing sees the mechanical and electromagnetic layers laminated prior to composite curing. The layers are then fit into a silicone mold and cured {\it together}: hence ``co-cure." When curing is complete, the entire structure exists as a single object and does not de-laminate. Co-curing offers a number of advantages that are also discussed in depth in Table \ref{tab:cocure_advantages}. In particular, 
\begin{itemize}
    \item Aligning the antenna sheet and raw composite fiber weave layer when flat, rather than when shaped, is simpler, more accurate, and lends itself to industrial scalability much more than the alternative.
    \item Eliminating the need for an additional adhesive layer makes the design lighter, more consistent, and simplifies the electromagnetic environment around the antenna which serves to improve performance.
\end{itemize}

\begin{table*}[t]
  \centering
  \begin{tabular}{|l|m{7cm}|m{7cm}|}
  \hline
{\bf Performance Metric} & {\bf Adhere After Curing} & {\bf Co-Curing}\\
        \hline
Mass & Extra adhesive adds mass. & We only use existing epoxy from the substrate which is impregnated using the minimum mass necessary to achieve mechanical integrity.\\
\hline
Mass variation & Spray adhesives may add spatial non-uniformity due to sputtering; tapes introduce mass variation owing to an unknown and possibly high-variation manufacturing process. & Epoxy impregnation process is highly uniform and curing under pressure insures epoxy flows through structure evenly.\\
\hline
EM losses & Additional adhesive would compound losses beyond those attributed to composite epoxy. The (limited) adhesive options have poorly documented EM properties and performance is difficult to model. & The composites community is rich, diverse, and has led to diverse epoxy options with well-documented EM properties. Minimal epoxy lends itself to minimal losses.\\
\hline
EM variation & Inconsistent adhesive application leads to inconsistent EM performance. If losses are nonuniform, antenna pattern can suffer. & Consistent epoxy layer is electromagnetically consistent. Performance can be well-modeled.\\
\hline
Alignment accuracy & Alignment involves more degrees of freedom and accuracy cannot benefit from traditional alignment practices and equipment. & Alignment process in 2D has only two degrees of freedom and can benefit from industry standards practices like fiducial usage.\\
\hline
Manufacturing simplicity & Adhering in 3D involves another step that is potentially complex. & No additional manufacturing steps above those required to make the cured composite.\\
\hline
Manufacturing scalability & Complex adhering lends itself poorly to scalability. & Adhering 2D layers is a solved problem and the additional conductive layer should not impact scalability.\\
\hline
Mechanical strength & Adhesive sprays and tapes are not designed for strength or flexibility; their mechanical performance is also more difficult to model. & Cured epoxy is designed for (well-modeled) mechanical strength.\\
\hline
Thermal durability & Potential coefficient of thermal expansion (CTE) mismatch between adhesive and conductive layer or with adhesive and composite. Adherence may be temperature-dependent. & No additional material reduces potential for CTE mismatch. Epoxy can be selected for thermal durability and CTE match with conductive layer.\\
\hline
\end{tabular}
  \caption{Advantages of co-curing the electronic layer to structural layer}
  \label{tab:cocure_advantages}
\end{table*}

The utility of co-curing extends beyond this specific radiator and can serve as a new platform for making flexible or non-planar electronics. By etching the electronics and then curing to a composite, we have the freedom to reshape the electronics into an endless array of shapes and to tune the physical properties - flexibility, strength, mass, thickness - as the application demands. The process of co-curing is essentially a platform for next-generation electronic devices and can see applications in conformal array design, human-machine interfaces, robotics, aeronautics, and other diverse fields with applications that demand electronic-mechanical co-design.

\subsection{Electromagnetic Design}

\begin{figure}[t]
\centering
    \includegraphics[width=0.48\textwidth]{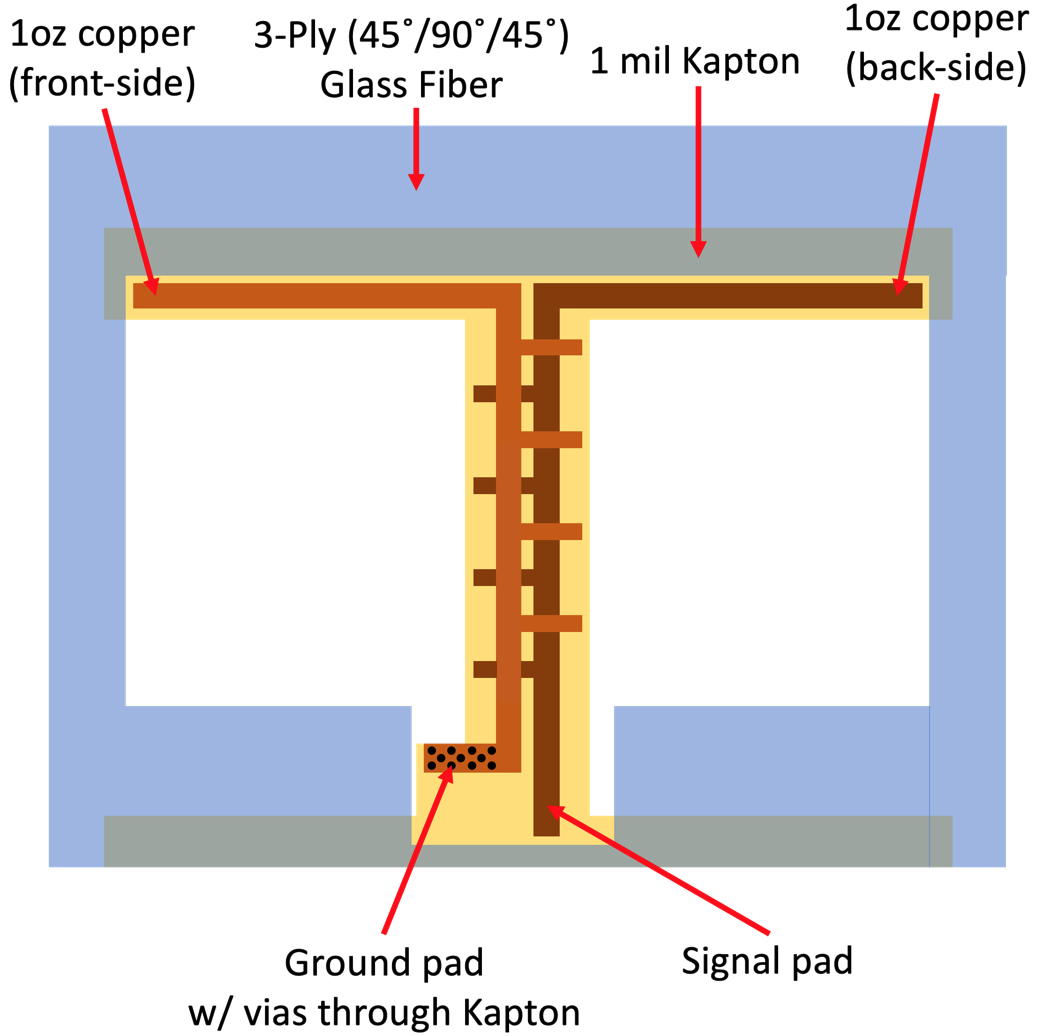}
    \caption{Front (Flattened) Diagram of Flexible Dipole Antenna. Not to scale.}
    \label{fig:flat_dipole_diagram}
\end{figure}
\begin{figure}[t]
\centering
    \includegraphics[width=0.48\textwidth]{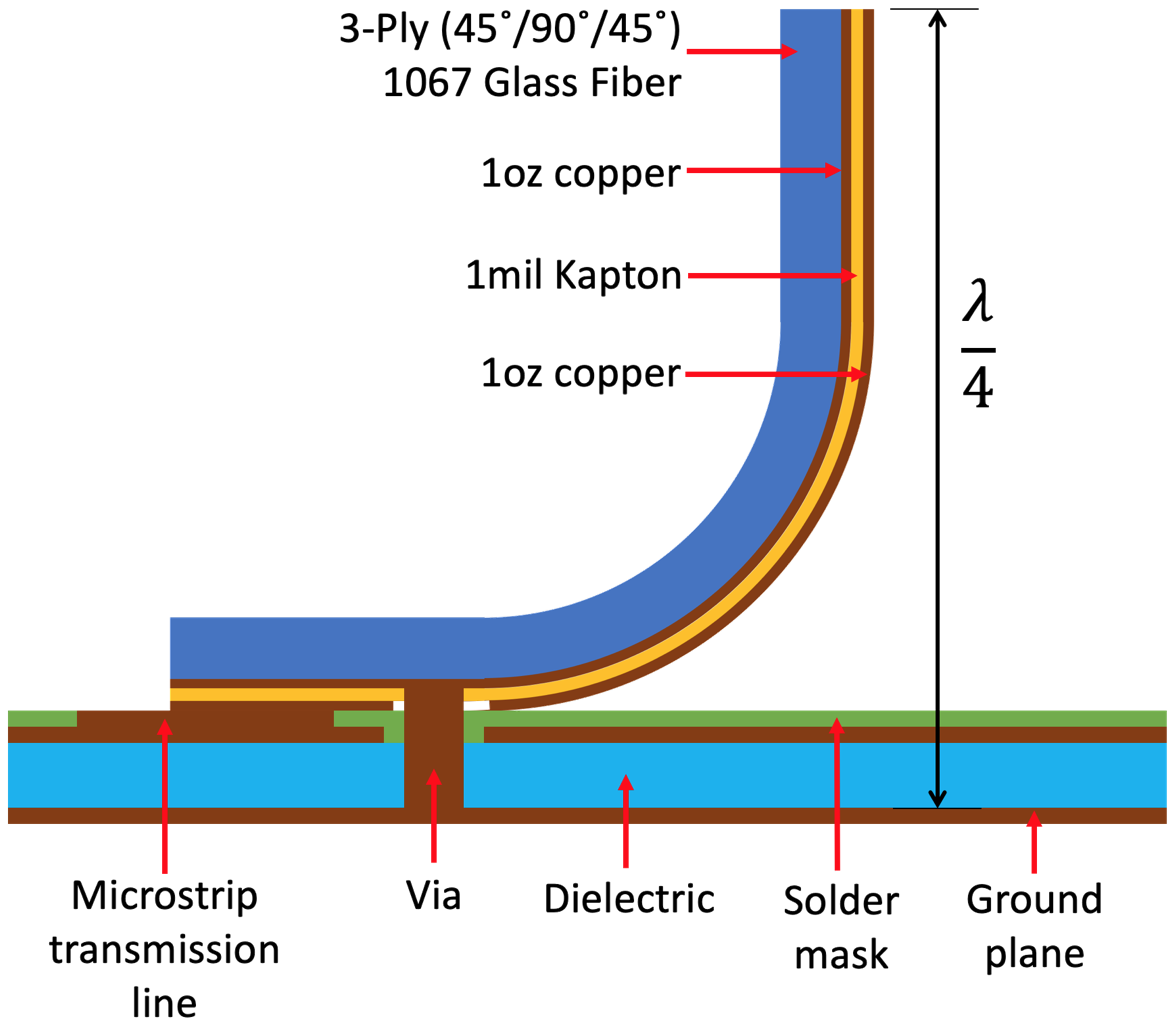}
    \caption{Side Diagram of Flexible Dipole Antenna. Not to scale.}
    \label{fig:side_dipole_diagram}
\end{figure}

The presented antenna, shown in Fig. \ref{fig:flat_dipole_diagram} and Fig. \ref{fig:side_dipole_diagram}, is composed of three sub-components: a circuit board contact, a feed transmission line, and two radiating arms. The antenna is driven by a single-ended microstrip transmission line on the PCB, with the radiator ground pad connecting to the transmission line ground and the radiator signal pad connecting to the microstrip line trace.

The feed transmission line, which rises in a "J" shape from the board, is critical to proper functioning of the antenna as it must accomplish single-ended to differential conversion (operate as a balun) and impedance matching between $50\Omega$ at the transmission line to $73 + 42.5j\Omega$ at the dipole arms. To achieve a near $50\Omega$ impedance on the thin and narrow polyimide substrate of the feed, the design relies on distributed capacitance between the two feed lines.

\begin{figure}[t]
\centering
    \includegraphics[width=0.48\textwidth]{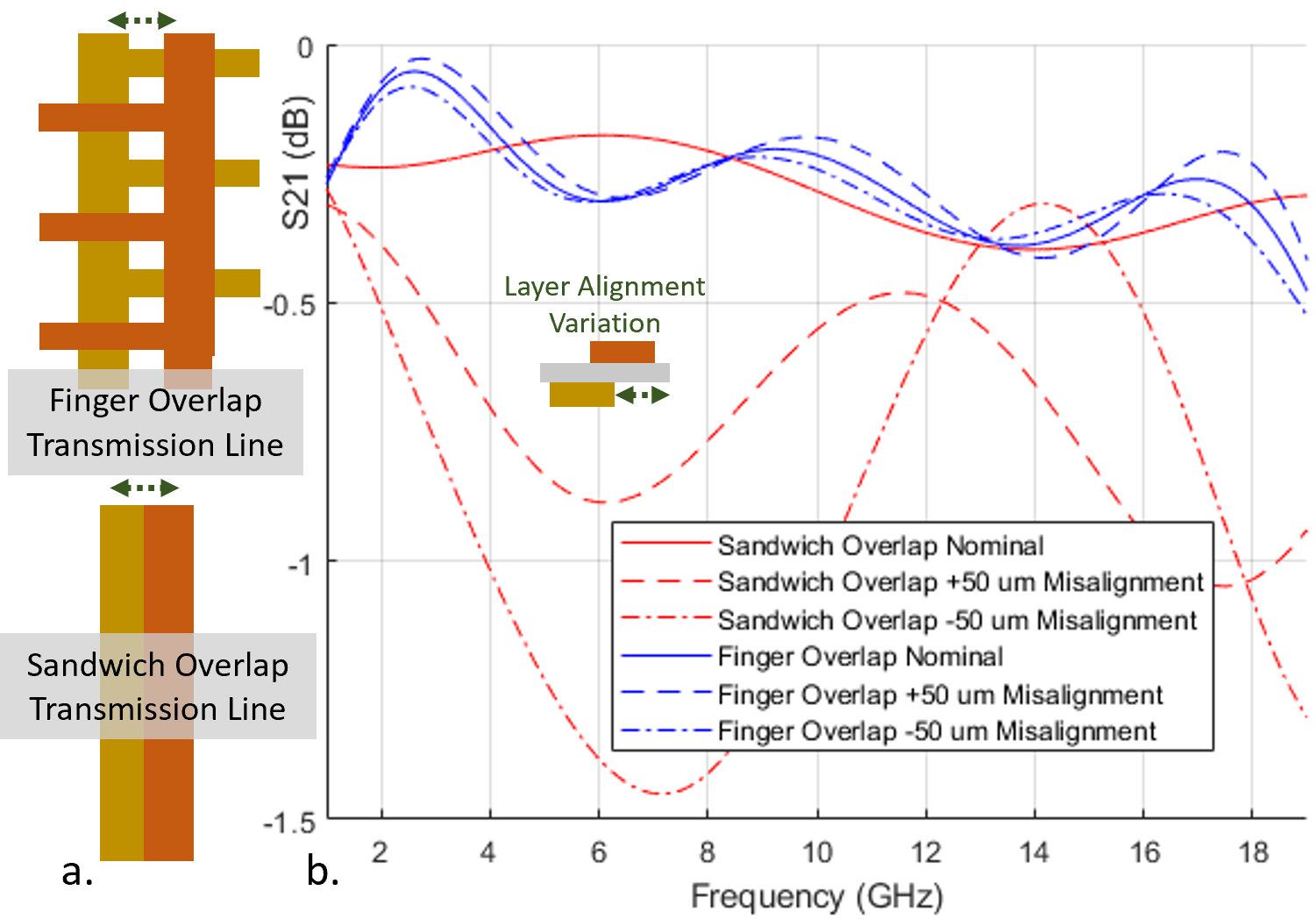}
    \caption{a. Sandwich and finger overlap feed transmission line designs b. FDTD simulation of s-parameters of the two transmission line designs. $\pm 50\mu m$ alignment error is added from the nominal design dimensions. Reproduced from \cite{APSURSI_FIKES} with permission.}
    \label{fig:misalignment}
\end{figure}

The simplest way to achieve this capacitance is using an ``edge overlapping sandwich” design as depicted in Fig. \ref{fig:misalignment}a. However, the relative position of etched copper on either side of the polyimide is subject to significant manufacturing variation and, thus, capacitance variation, owing to the (relatively) inaccurate layer alignment process. Variation in capacitance can greatly degrade matching and, thus, the antenna efficiency. High manufacturing sensitivity is undesirable, thus motivating use of the ``finger" overlapping design conceived of for this work. Fig. \ref{fig:misalignment} illustrates the effect of manufacturing layer misalignment on the transmission ($|S_{21}|$) through the sandwich and finger overlap transmission line designs. By using fingers to achieve the necessary distributed capacitance for the feed transmission lines, the effect of $\pm 50\mu m$ alignment errors from manufacturing on $|S_{21}|$ is reduced from \textgreater 1.2 dB to \textless 0.2 dB.

\section{Manufacturing}

\begin{figure*}[t]
\centering
    \includegraphics[width=\textwidth]{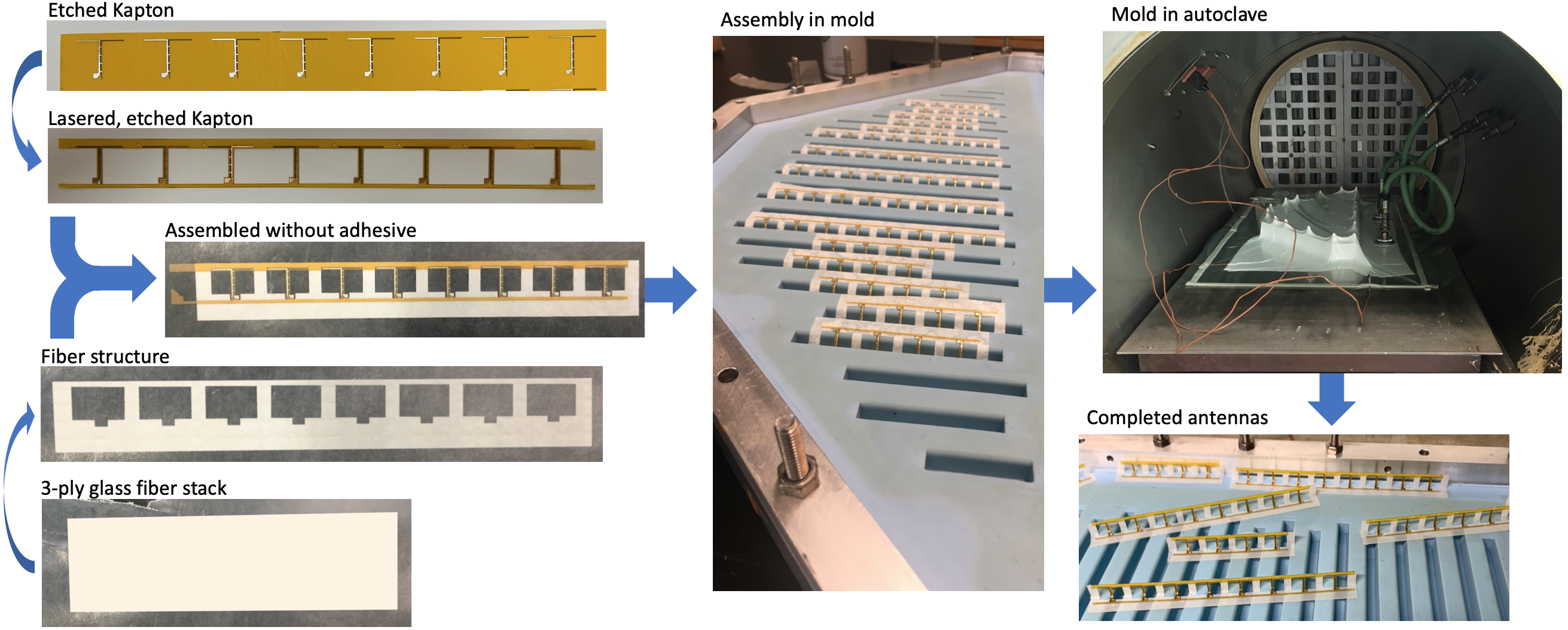}
    \caption{Manufacturing Flow Chart}
    \label{fig:manufacturing_flow_chart}
\end{figure*}

The manufacturing process, shown as a flow chart in Fig. \ref{fig:manufacturing_flow_chart}, begins with two processes that can happen in parallel: etching and cutting of the conductive layer and stacking and cutting of the laminate. These two layers are subsequently aligned and fitted into a silicone mold. The mold is vacuum-sealed and put into an autoclave. The layers are cured at temperature and pressure for 2 hours to form antennas. These steps are discussed in depth below:

\subsection{Conductive Layer}

The conductive layer is a 2-layer polyimide printed circuit board with sections of polyimide removed to reduce mass and electromagnetic loss. The board begins as raw PCB clad (1 mil polyimide, 0.5oz copper on either side), which is first drilled to produce vias that are used on the antenna pads. The boards are plated\footnote{Once plated, the metal layer becomes 1oz thick on both sides.}, patterned, and then etched on both sides. To reduce mass, we use a laser to cut out ``windows" of unused polyimide below the dipole arms. This also slightly reduces electromagnetic radiation losses. These sheets are now complete and ready to be stacked on top of the laminate.

\subsection{Laminate}

The laminate begins as rolls of single-layer, plain-weave 1067 glass fiber impregnated with a matrix of Patz F-4 resin. The glass laminate is cut to size and stacked in a symmetric 45$^\circ$/90$^\circ$/45$^\circ$ configuration which gives us the desired balance of toughness and stiffness. The layers are lightly laminated using a low-temperature flat iron. The gentle heat from the iron causes the epoxy in the laminates to flow and helps the layers stick together. The stack is then cut using a ZUND m-1600 CNC machine to remove excess width and produce the desired shape.

\subsection{Alignment and Curing}

The two layers are then ready for assembly. Small holes on the conductive layer that line up with corners on the laminate layer act as fiducials. This ensures that everything is aligned accurately such that the laminate does not cover any metal regions on the conductive layer. Once aligned, the two layers are adhered through brief contact with a clothing iron set to a light heat setting.

Then, the full stack is fitted into a custom silicone mold with the desired shape \cite{silicone_mold}. The top half of the mold presses the stack into the antenna shape; the entire mold is covered with a fabric mesh\footnote{A fabric mesh, like loose cotton or another highly porous fabric, allows air to flow between the mold and the vacuum bag, ensuring a very high quality vacuum seal.} and a vacuum bag before being vacuum sealed. The vacuum-sealed mold is placed into an autoclave with internal vacuum tubes; the vacuum tubes are connected to valves on the vacuum bag to remove any trapped air. The vacuum is maintained during curing.

The autoclave is sealed and curing takes place under a set temperature and pressure profile. The profile sees the temperature rising slowly from room temperature until it reaches 120$^\circ$C. The temperature is held at 120$^\circ$C for 2 hours and then brought down to room temperature. An oven pressure of 80 psi is applied for the entire cure time.

\section{Electromagnetic Performance}

\begin{figure}[t]
\centering
    \includegraphics[width=0.48\textwidth]{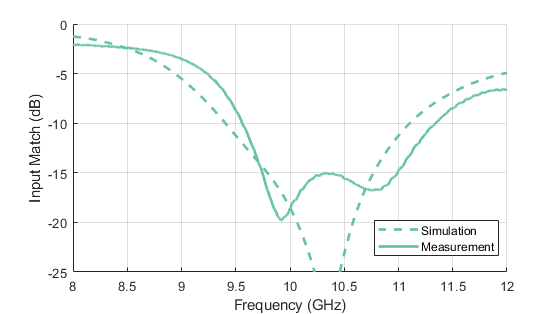}
    \caption{Simulated and measured (sample 4) antenna input match. Reproduced from \cite{APSURSI_FIKES} with permission.}
    \label{fig:Match}
\end{figure}

The array performance was characterized using an FDTD simulation of the elements in an 8x1 linear array. The simulation used open boundary conditions and modeled the entirety of the element design, including the feed line, balon, dipole arms, and glass fiber backing. As we can see in Fig. \ref{fig:Match}, the antenna's resonance peak is at 10.328 GHz, with a better than -10dB return loss over a bandwidth of 1.7 GHz (16\%) from 9.4-11.1 GHz.

The array performance was also characterized using measurement of a ``mock" 1x8 antenna array with identical dimensions and design as the simulation. The antenna's match ($S_{11}$) was measured while adjacent antennas were terminated to a (matched) $50\Omega$ load. Fig. \ref{fig:Match} shows the measured match for ``Element 4" plotted alongside the simulation results.

\begin{figure}[t]
\centering
    \includegraphics[width=0.48\textwidth]{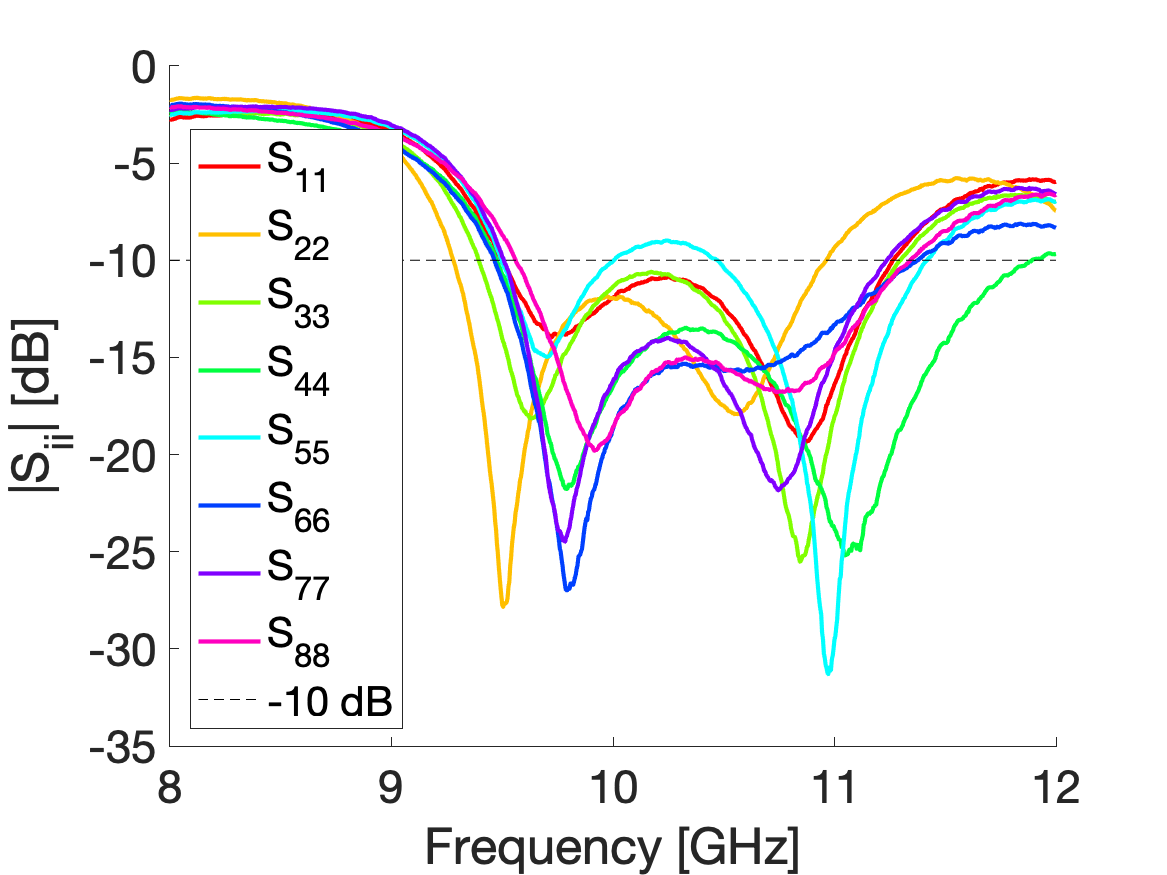}
    \caption{Manufacturing consistency of eight array elements.}
    \label{fig:match_8}
\end{figure}

\begin{figure}[t]
\centering
    \includegraphics[width=0.48\textwidth]{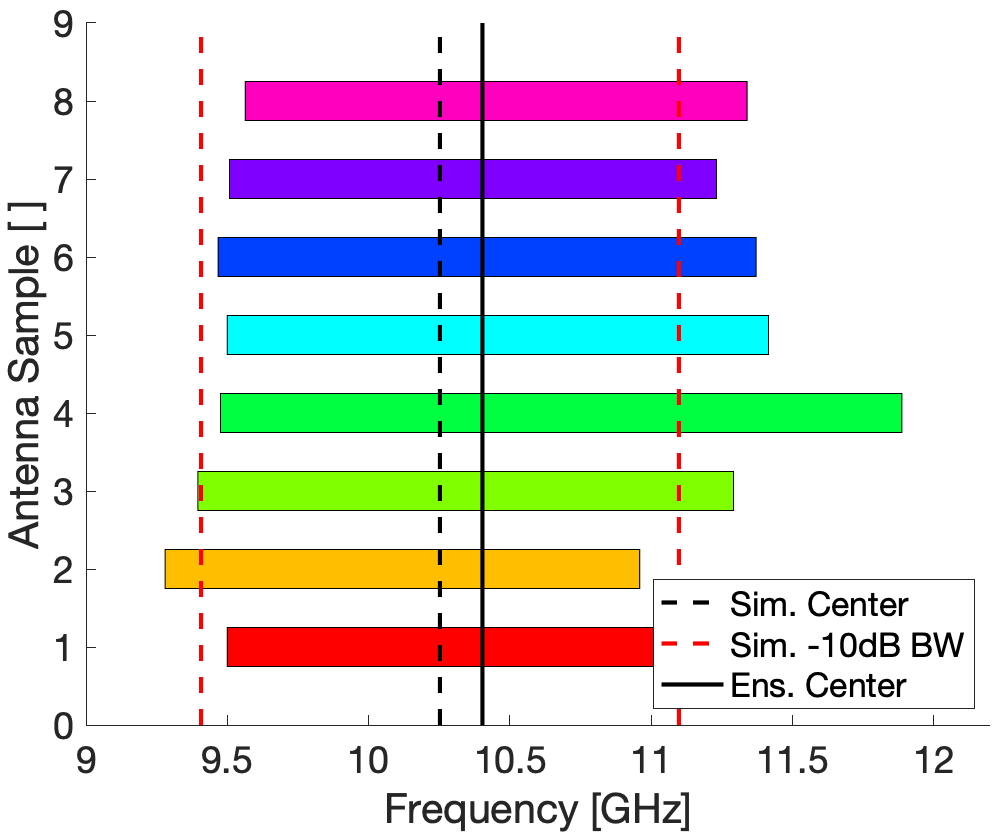}
    \caption{Bandwidth consistency of eight array elements. Rectangles show the range of $f$ for which $S_{11}(f) < -10$dB for each element.}
    \label{fig:BW_8}
\end{figure}

Moreover, to demonstrate the consistency of the manufacturing process, all 8 antennas were characterized; individual $S_{11}$ curves for all 8 elements are plotted together in Fig. \ref{fig:match_8}. To best demonstrate the shared overlap of all 8 elements' bandwidth, the lower and upper limit of the frequency range with a better than $-10$ dB return ratio was plotted for each element, shown in Fig. \ref{fig:BW_8}. The centers of the lower and upper limits for each antenna were averaged; this quantity is called the ``Ensemble Center," and is plotted and labeled on Fig. \ref{fig:match_8}. The ``Ensemble Center" (10.4032 GHz) is close to the simulated center (10.253 GHz). The up-tuning of the ``Ensemble Center" relative to simulation can be attributed to a higher {\it measured} upper limit (11.345 GHz compared to 11.099 GHz in simulation); the measured lower limit (9.462 GHz) is close to simulation (9.4073 GHz).

\begin{figure}[t]
\centering
    \includegraphics[width=0.48\textwidth]{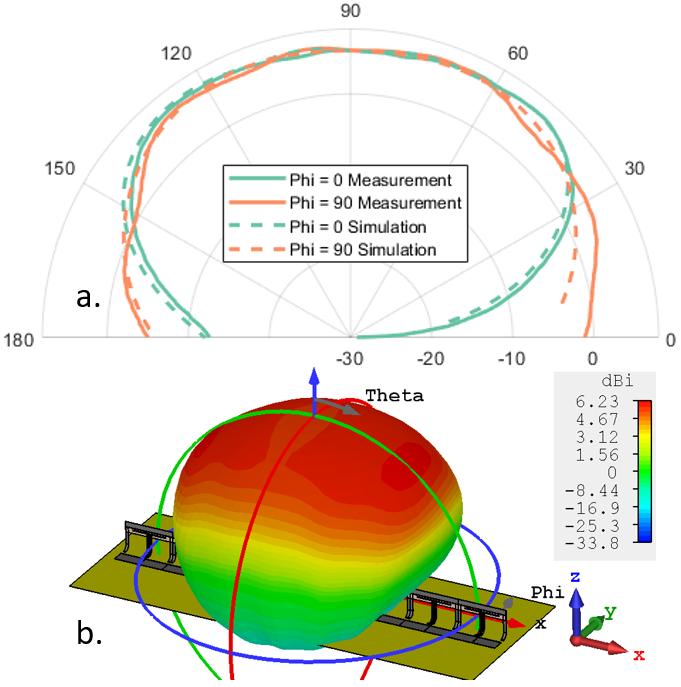}
    \caption{a. Simulated and measured, collapsible pop-up dipole element patterns. b. Simulated 3D pattern with scale and axis. Reproduced from \cite{APSURSI_FIKES} with permission.}
    \label{fig:Pattern}
\end{figure}

The simulated and measured gain patterns for orthogonal cuts are presented in Fig. \ref{fig:Pattern}. The measured and simulated broadside gain match closely - both are 5.3 dBi. As we can see, both measured and simulated patterns exhibit slight lobe splitting along the array axis. The antenna demonstrates a half-power beam width close to 110$^{\circ}$ in both the $\phi=0^{\circ}$ and $\phi=90^{\circ}$ cuts.

\section{Mechanical Performance}
Space missions involve extreme conditions (vibration, temperature, radiation) that place demanding requirements on hardware well beyond those introduced in ground operation. It is typical, and often required, to perform a smattering of testing to ensure parts do not degrade or malfunction in these conditions. Many of the tests involving extreme thermal and mechanical environments were performed on this array\footnote{Radiation testing was not performed on this part because it has no active electronic or otherwise vulnerable parts.}:
\begin{itemize}
    \item The array was shake-tested in 3 dimensions using the spectrum from SpaceX's ``Rideshare Payload User's Guide" \cite{spacex_falcon} to ensure the parts would physically survive launch.
    \item The antennas' match and coupling was measured during a thermal sweep in an absorptive thermal chamber to ensure operation at the expected temperature range and limited thermal sensitivity.
    \item The array was subject to several days of high frequency thermal cycles covering most of the expected temperature range to ensure that it could survive extreme temperature oscillations associated with orbiting in and out of sunlight.
    \item The array was immersed in liquid nitrogen repeatedly to ensure its survival to rapid temperature drops and increases.
    \item The array was subjected to long periods of time in the flattened configuration at elevated temperature to simulate months or years of stowage.
\end{itemize}

\subsection{Vibration}

Critical to any space mission's success is the ability to withstand the spectrum of mechanical vibrations associated with a launch. These vibrations can often be wideband and very high amplitude. We performed a set of shake tests at and above mission qualification levels to confirm that our array can survive these conditions, deployed or stowed (flat).

Shake testing for mission qualification included four classes of tests:
\begin{enumerate}
    \item Sine Vibe - A frequency-domain sweep in accordance with the rocket's spectrum.
    \item Random Vibe - A time-domain test which generates a random vibration waveform in accordance with the rocket's spectrum.
    \item Sine Burst - A time-domain test at a single (low) frequency meant to test high displacement vibration.
    \item Shock - A time-domain test meant to simulate sudden accelerations associated with dropping or banging the component.
\end{enumerate}
These tests are performed in succession at various levels. After each test, the component is checked visually and physically to determine the presence of damage or mechanical change. In addition, a low-amplitude frequency sweep (sine sweep) is performed before and after each vibration test to detect changes in the object's mechanical spectrum. The spectrum is essentially a mechanical fingerprint of the object; significant changes to the spectrum indicate damage or small changes (a screw loose, a displaced object, etc.) to the payload.

\begin{figure}[t]
\centering
    \includegraphics[width=0.48\textwidth]{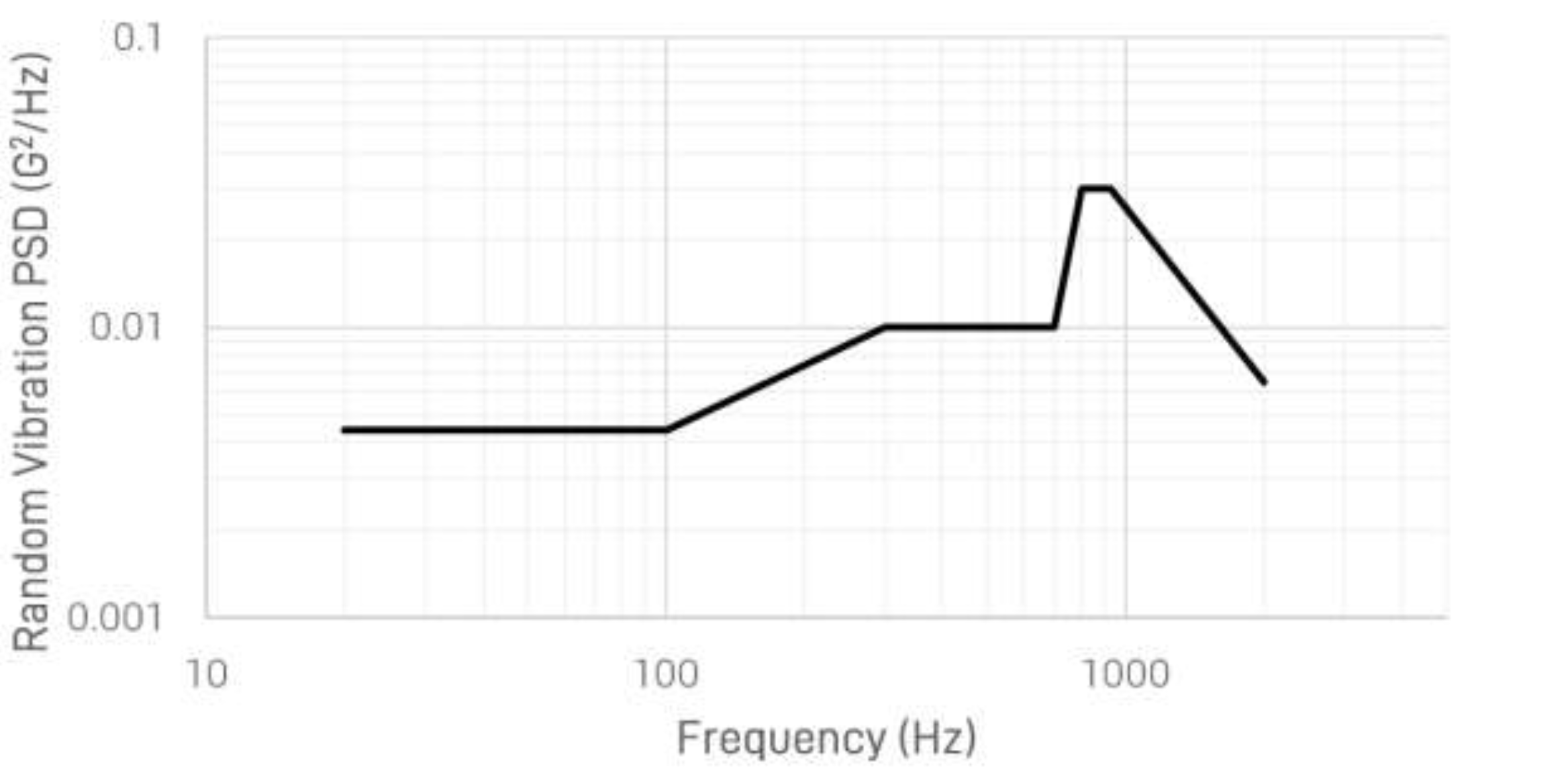}
    \caption{SpaceX Rideshare User Guide random vibration spectrum, adapted from \cite{spacex_falcon}.}
    \label{fig:array_mechanical_spectrum}
\end{figure}

Our array was tested in compliance with SpaceX Rideshare Payload requirements for sine vibration, shock, and random vibration \cite{spacex_falcon} in anticipation of space deployment. The spectrum profile for random vibe testing is shown in Fig. \ref{fig:array_mechanical_spectrum}.

The antenna array presented no significant and/or relevant mechanical resonance peaks and suffered no damage throughout the entire testing process. The array's mechanical spectrum was stable throughout the testing process, in spite of being subject to vibration amplitudes 25\% in excess of the SpaceX launch profile.

\subsection{Thermal Limits}

\begin{figure}[t]
\centering
    \includegraphics[width=0.48\textwidth]{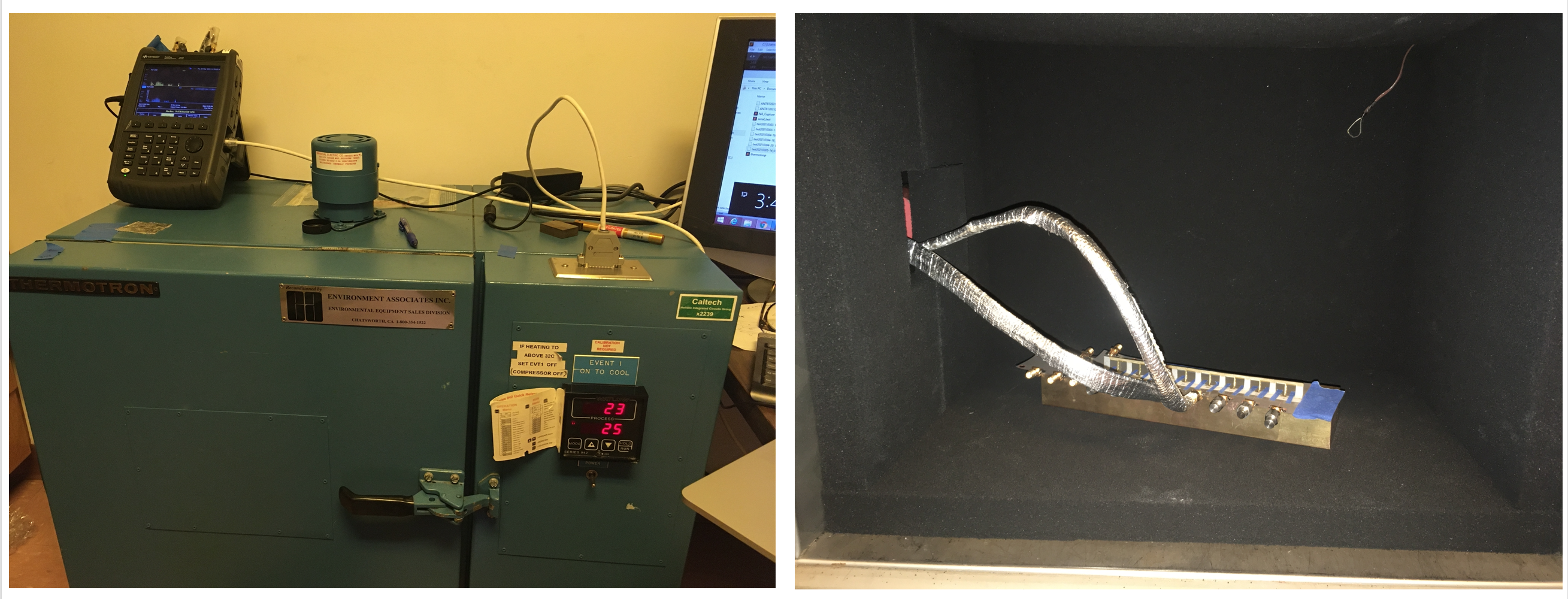}
    \caption{Thermal Sweep Experimental Setup. Left: Setup showing thermal chamber, vector network analyzer, and PC. Right: Inside of thermal chamber showing array on top of large brass bar.}
    \label{fig:thermal_sweep_setup}
\end{figure}

To understand the antenna's thermal limits and characterize temperature-dependent changes in the antenna's performance, we performed a slow temperature sweep in a thermal chamber while measuring the antenna's match and mutual coupling. 

\subsubsection{Thermal Limit Values}
The range of temperatures tested was based on SpaceX's Rideshare User Guide \cite{spacex_falcon}. The document projects expected payload temperatures between [-20, 52]$^\circ$C, with qualification testing to be done at $\pm 10^\circ$C. We wanted to further exceed qualification testing limits, and used a temperature range of [-45, 80]$^\circ$C. The maximum of $80^\circ$C is slightly less than the $92^\circ$C maximum that SpaceX projects for the fairing temperature at the peak of flight.

\begin{figure}[t]
\centering
    \includegraphics[width=0.48\textwidth]{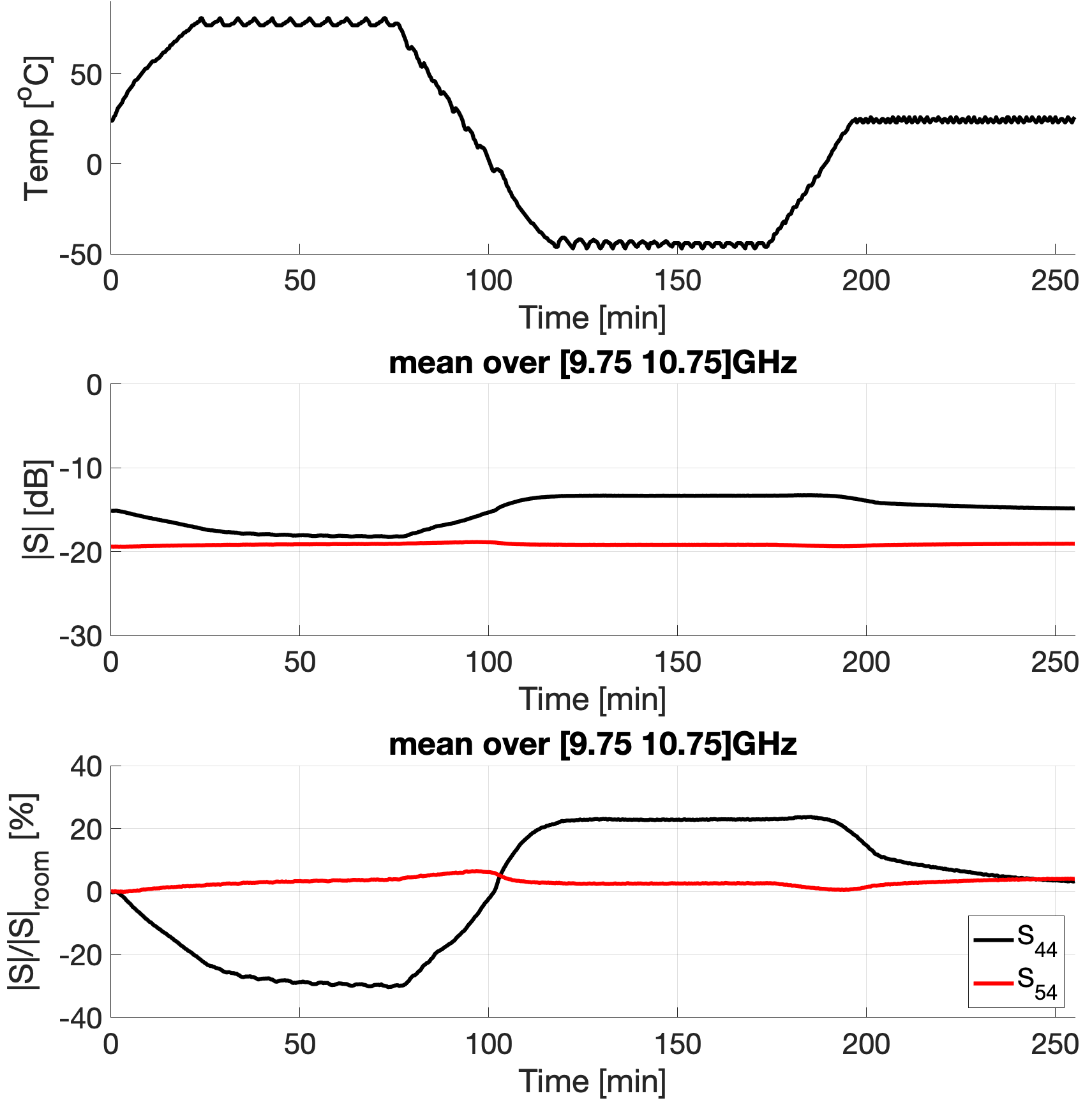}
    \caption{Top: Thermal sweep temperature profile. Middle: Scattering parameter over [9.75 10.75] GHz. Bottom: Percent change of average scattering parameter relative to room temp.}
    \label{fig:thermal_sweep_temperature}
\end{figure}

\begin{figure*}[t]
\centering
    \includegraphics[width=\textwidth]{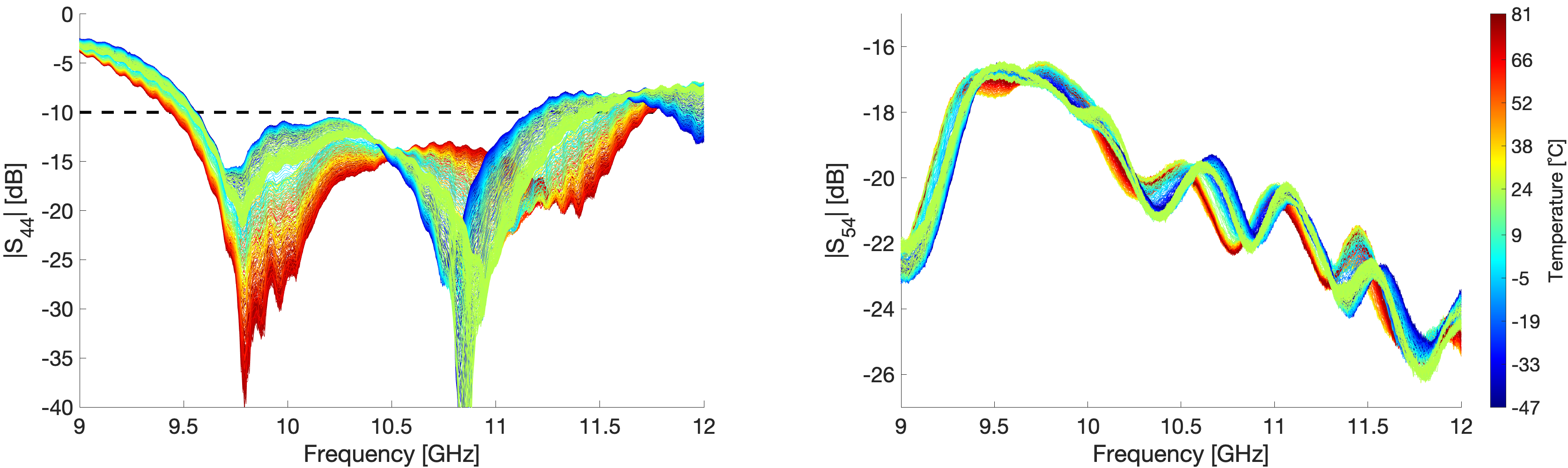}
    \caption{Left: Match spectrum over temperature sweep. Right: Mutual coupling over temperature sweep.}
    \label{fig:thermal_sweep_S11_S21}
\end{figure*}

\begin{figure}[t]
\centering
    \includegraphics[width=0.48\textwidth]{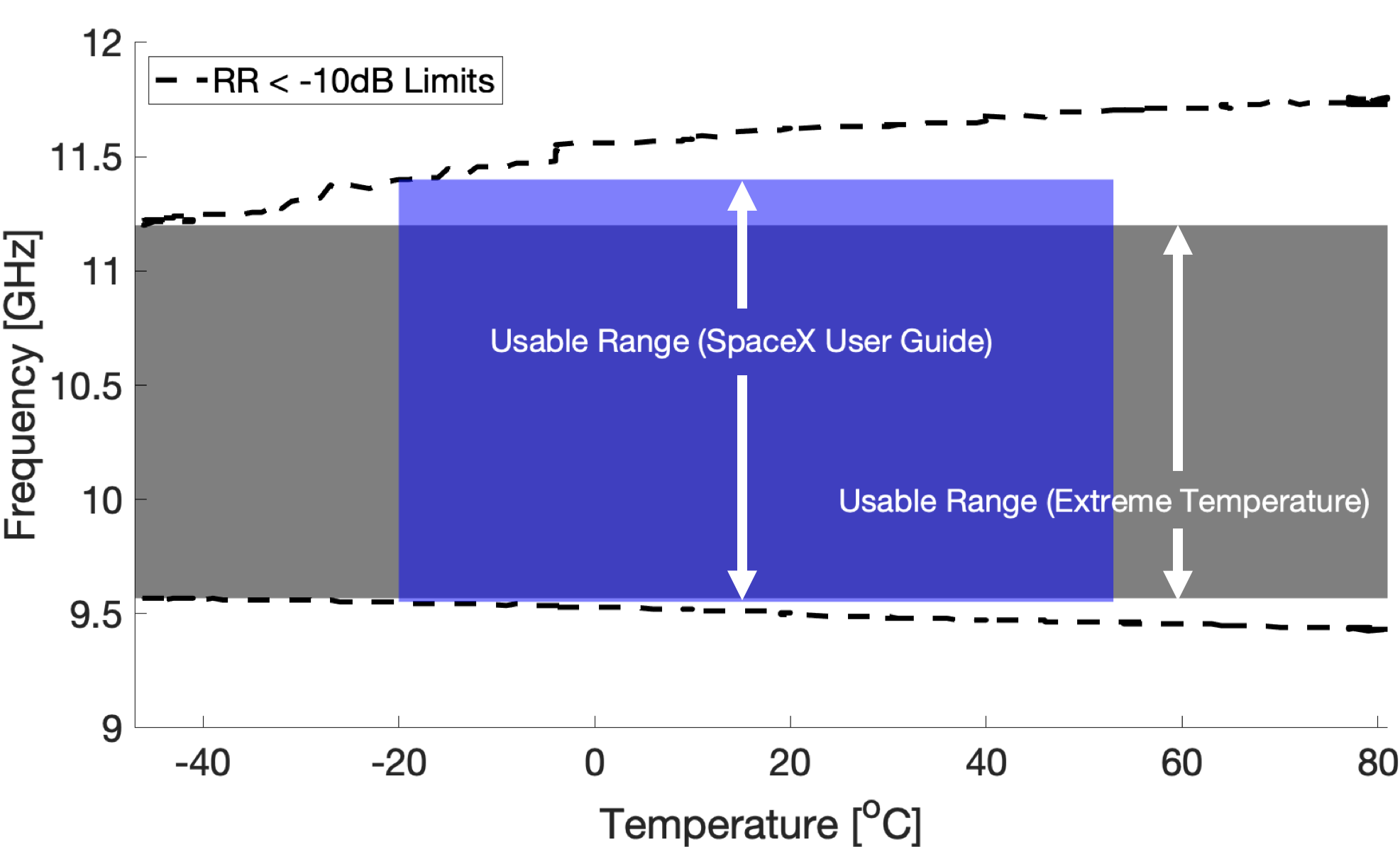}
    \caption{The antenna's ``usable range" based on different thermal limit guidelines. In dashed lines are the upper and lower limits of the region in which the antenna has a return ratio (RR) $< -10$dB. Based on SpaceX's Rideshare User Guide's projected temperature range, our radiator has an effective bandwidth of 1.848GHz, as compared to 1.632GHz using a much more extreme temperature range.}
    \label{fig:usable_range}
\end{figure}

\subsubsection{Test Setup}
A thermal chamber was lined on (all) 6 internal walls with RF absorbing foam specifically chosen to have low reflection characteristics. The array was placed inside the chamber and SMA cables were fed to the two central antennae (\#4 and \#5) while the others were terminated with 50$\Omega$ loads. The array was mounted on a thick brass bar to ensure antennas are consistently being held at the programmed temperature. A thermocouple was mounted to the bar to ensure the bar was at the programmed temperature. The SMA cables and the thermocouple cable were fed through a small hole on the side of the chamber. The hole was plugged up with a round of thermally-insulating foam. This setup is shown in Fig. \ref{fig:thermal_sweep_setup}.

While in the chamber, the array was subjected to the thermal sweep according to the following schedule:
\begin{enumerate}
\item Start at room temperature (24$^\circ$C).
\item Ramp up to 80$^\circ$C at 2.4$^\circ$C/minute.
\item Soak (hold) for 50 minutes.
\item Ramp down to -45$^\circ$C at 3.0$^\circ$C/minute
\item Soak (hold) for 50 minutes.
\item Ramp up to 24$^\circ$C at 2.4$^\circ$C/minute.
\item Stop.
\end{enumerate}

\subsubsection{Results}
The chamber temperature during the sweep, shown in Fig. \ref{fig:thermal_sweep_temperature}, closely matches the desired, programmed profile. Low amplitude and low frequency temperature cycling presents during hold (``soak") periods, owing to the chamber's extremely simple (bang-bang) feedback system using either a heater and a cooler. Oscillations have a period of $\approx 5$ min and temperature varies with a 2 degree amplitude ($T \in [77, 81]^\circ C$).

Fig. \ref{fig:thermal_sweep_S11_S21} presents the match ($S_{44}$) and mutual coupling ($S_{54}$) spectra over the temperature sweep. In this plot, each curve is a sampled spectrum, color-coded by the temperature of the element at the time of the sweep.

As we can see, high temperatures improved the antenna match over most of the bandwidth and slightly improved coupling; low temperatures had the opposite effect. The ``usable" range of frequencies is defined as the range of frequencies for which the return ratio is less than -10dB for {\it all} possible temperatures. This range is a function of the temperature range chosen and is visualized in \ref{fig:usable_range} and in the table below:
\begin{table}[h]
    \centering
    \begin{tabular}{r|c|c|c|c|c}
        Temperature Spec & $T_\text{min}$ & $T_\text{max}$ & $f_\text{min}$ & $f_\text{max}$ & Bandwidth\\
        & [$^\circ$C]  & [$^\circ$C] & [GHz] & [GHz] & [GHz]\\\hline
        SpaceX User Guide    & -20 & 52 & 9.55 & 11.40 & 1.848\\
        Extreme Temperatures & -45 & 80 & 9.57 & 11.20 & 1.632
    \end{tabular}
    \caption{Frequency ranges with less than -10dB return ratio for all temperatures within expectation, according to various temperature specifications.}
    \label{tab:my_label}
\end{table}

\begin{figure*}[t]
\centering
    \includegraphics[width=\textwidth]{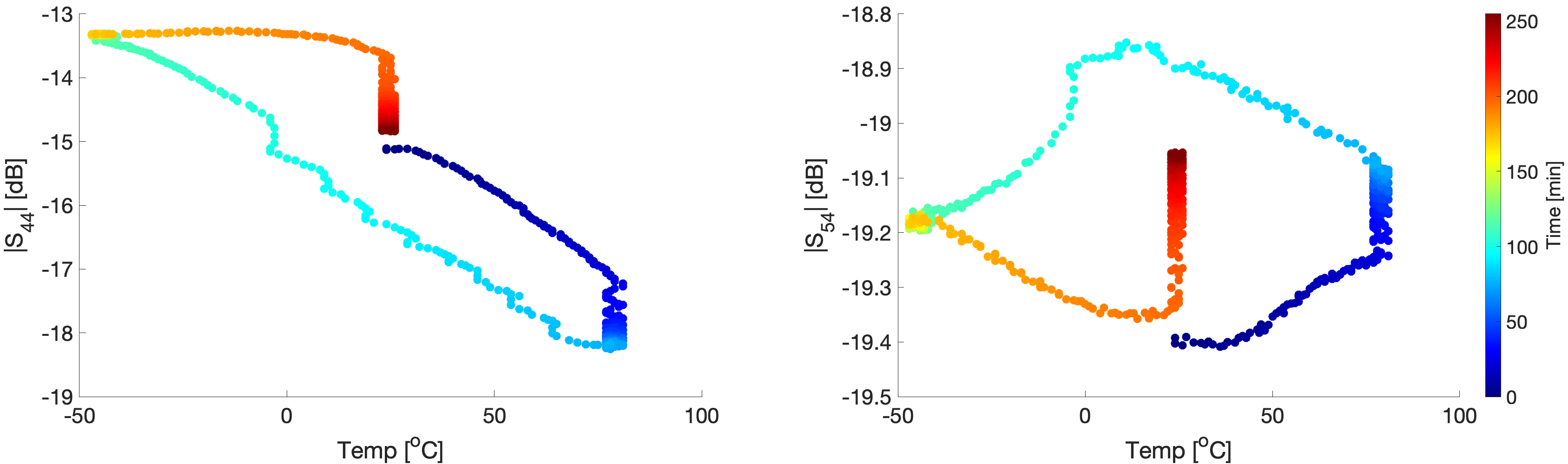}
    \caption{$\langle |S_{44}| \rangle_f$ and $\langle |S_{54}| \rangle_f$, the ensemble-averaged magnitude of the S parameters over $f\in [9.75, 10.75]$ GHz plotted against temperature; points are color-coded by time.}
    \label{fig:thermal_sweep_hysteresis}
\end{figure*}

However, it's difficult to draw conclusions about the array's temperature dependence from such a figure. To this end, we present $\langle |S_{44}| \rangle_f$ and $\langle |S_{54}| \rangle_f$ - the ensemble average of the magnitude of S parameters $S_{44}$ and $S_{54}$. The ensemble average is taken with respect to frequency, over the center of the array's bandwidth (9.75-10.75GHz). These figures are displayed as a function of both time and temperature in Fig. \ref{fig:thermal_sweep_hysteresis}.

As we can see, there is a roughly linear relationship between the magnitude of the S-parameter and temperature, with both coupling and match improving (going up and going down, respectively) as temperature increases during the first ramp.

However, the pattern is made more complex by apparent hysteresis in the antenna's behavior. Though parameters improve as temperature increases in the first ramp, they continue to improve during the hold period. When the ramp reverses, these effects also reverse, but at overall improved values (lower match and higher coupling). This is possibly due to a delay when the array reaches a set temperature as compared to the air temperature. This explanation is plausible given we mounted the array to a large brass block which has a large thermal time constant. Moreover, we can see what looks like settling in both the match and the coupling during the soak periods. Time-varying changes in match and transmission when the air temperature is held roughly constant implies the array may be ``coming up to" or ``coming down from" temperature during this time. This may also explain the disparity in the starting and ending $S$ parameters, despite the air temperature being equal for both.

The overall change in S-parameter value, expressed as a percent change relative to the starting (room temperature) value is small. We calculated the following changes in the ensemble-averaged magnitude of S-parameters over the tested temperature range:
\begin{align}
    &\Delta \langle |S_{44}| \rangle_f \in [-30.36, 23.69]\%\\
    &\Delta \langle |S_{54}| \rangle_f \in [-0.07, 6.54]\%
\end{align}
Assuming perfect array symmetry, $S_{54}$ can be bijected to study temperature-driven changes in a single antenna element's transmission. Bijecting $S_{54}$ and then taking the ensemble average of the magnitude, we see a percent change of $\Delta \langle |S_{54}| \rangle_f \in [-0.04, 3.27]\%$ over the entire temperature sweep range. 

Moreover, measurements taken after testing demonstrated the thermal sweep caused no long-term damage to the array.

\subsection{Thermal Cycling}
Important for space-ready components is the ability to withstand the high-frequency thermal oscillations caused by orbiting around the Earth. Satellites in Low Earth Orbit (LEO) complete one full rotation around the Earth every 90-120 minutes, during which time exposure to the Sun changes dramatically. With no convection, radiation is responsible for the satellite's temperature; changes to Sun exposure imply potentially high-amplitude thermal oscillations.

Such high-amplitude, high-frequency thermal oscillations introduce stress that can lead to structural or other damage that may permanently alter or destroy the components' performance. We performed thermal cycling tests to verify that our array can withstand such oscillations.

\subsubsection{Test Setup}
\begin{figure}[t]
\centering
    \includegraphics[width=0.48\textwidth]{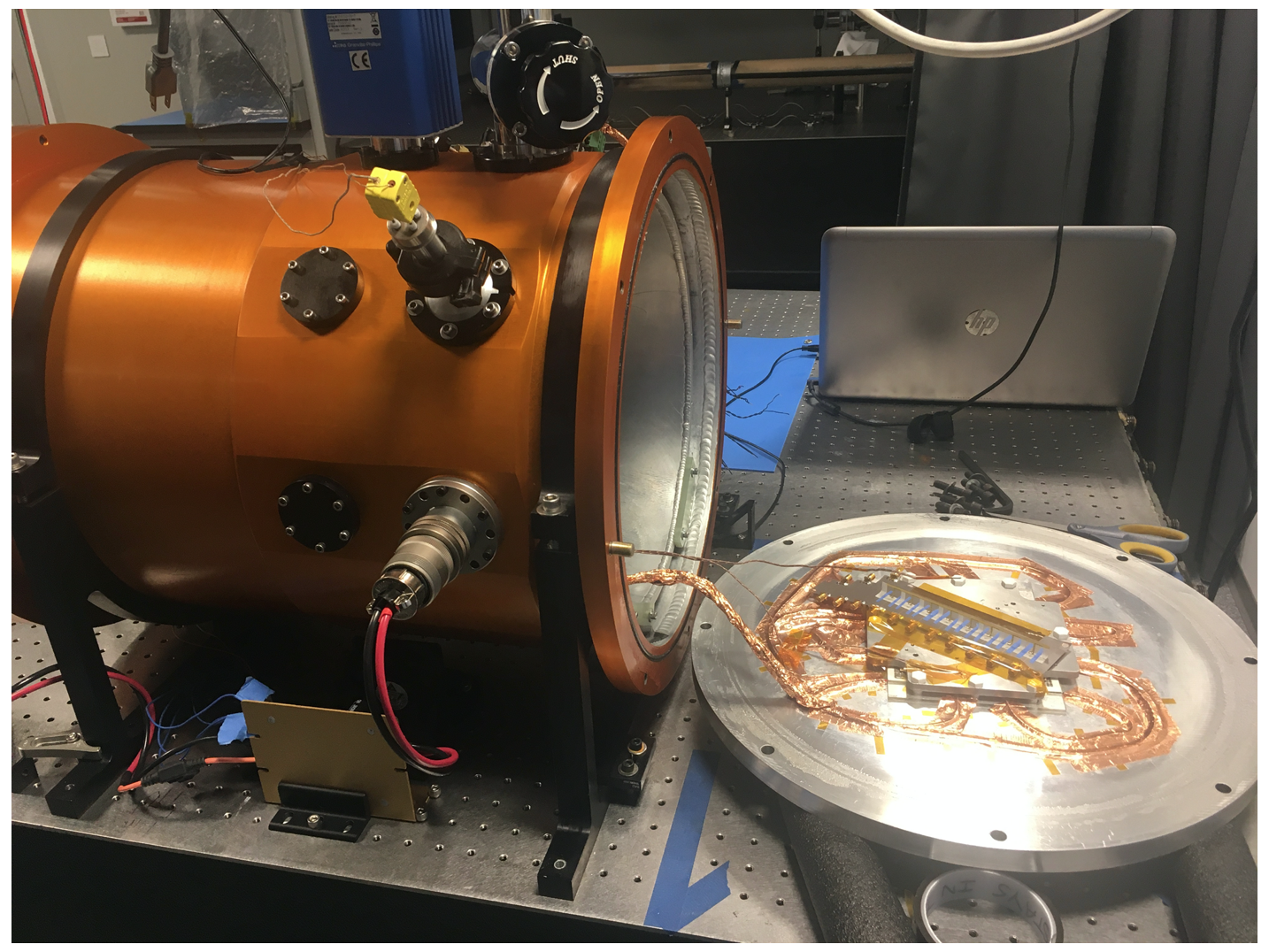}
    \caption{Thermal cycling vacuum chamber setup. Left: a large (orange) vacuum chamber. Right: our array mounted on a hot plate attached to the side wall of the chamber. Probes are attached to both the array and the hot plate.}
    \label{fig:thermal_cycling_setup}
\end{figure}

The tests were designed around two key goals:
\begin{enumerate}
    \item To replicate relevant space conditions (temperature range and vacuum).
    \item To maximize temperature ramp rate.
\end{enumerate}
The tests were performed in a thermal vacuum chamber to best accomplish the former goal. With regard to maximizing temperature ramp rate, the array was mounted directly on a hot plate. The low pressure within the chamber has the added effect of reducing the thermal time constant of the chamber - thus facilitating higher frequency oscillations - and reducing the potential for condensation at low temperatures. This setup is shown in Fig. \ref{fig:thermal_cycling_setup}.

\begin{figure}[t]
\centering
    \includegraphics[width=0.48\textwidth]{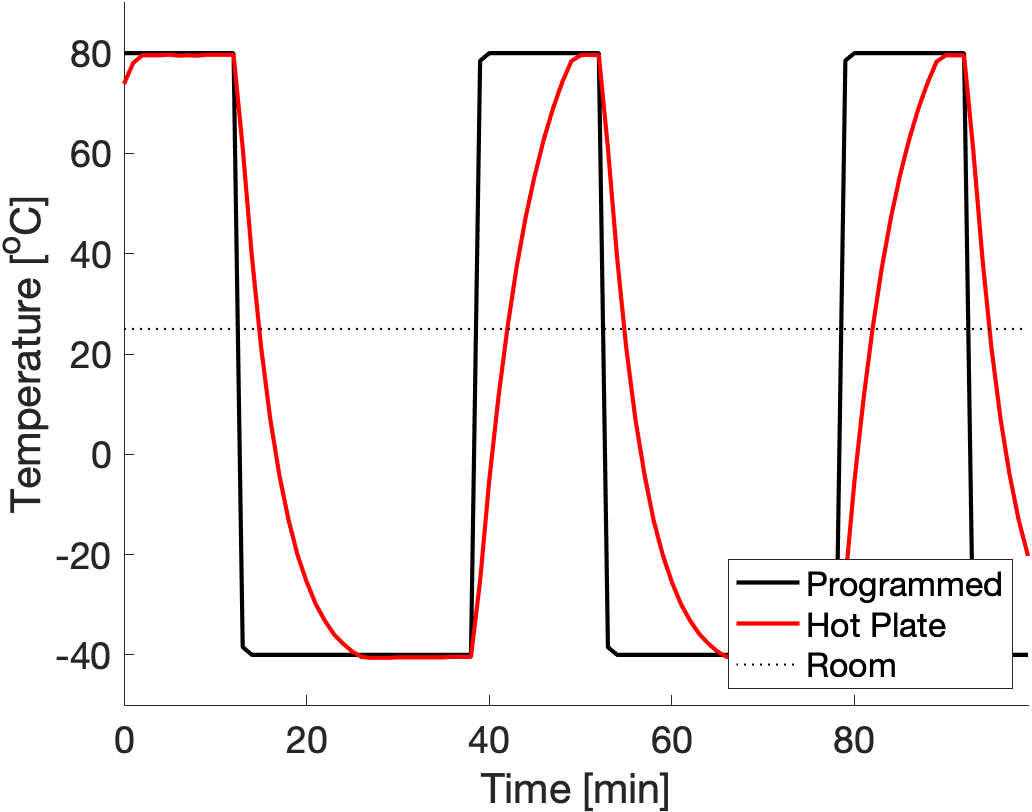}
    \caption{Thermal Cycling Profile showing programmed and plate temperatures.}
    \label{fig:cycling_profile}
\end{figure}

The vacuum chamber pressure was lowered to the chamber's minimum of $\approx 1\times 10^{-5}$ torr\footnote{This minimum was limited by leakage induced by imperfect (custom) sealant o-rings on the chamber end(s).}. Temperature probes were placed on the hot plate and on the array for remote monitoring of both temperatures during testing. Temperature limits were set based on the hot plate's minimum and maximum: $-40^\circ$C to $+80^\circ$C. At these plate temperatures, the array was only able to reach between $-26^\circ$C to $+65^\circ$C, even in steady-state conditions where the antenna had ample time to conduct heat from the plate. This range was cycled at the minimum period of 40 minutes during which both the plate and the array reached their full temperature range potential. These values were used to program an (asymmetric) square wave\footnote{The minimum temp is set for approximately 2/3 of the period and the maximum for the remaining 1/3. This asymmetry is due to observed differences in the settling times at the high and low temperatures. Most of the time, the hot plate reached the lowest temperature ``early" and the plate was held at the low temperature for the last 13 minutes of the cycle. Realistically, the cycle rate could have been higher while achieving the same temperature limits.} temperature profile, shown in Fig. \ref{fig:cycling_profile}. The profile was run 100 times back-to-back over 3.5 days.

\subsubsection{Results}
The temperature range and period corresponds to an average rate of $4.55^\circ$C/min for the array and a peak rate of $6.74^\circ$C/min\footnote{Peak rate reflects the calculation of temperature change during only that portion of the cycle when temperature is actually changing.}.

We can estimate peak thermal cycling rate ($r^\text{peak}_T$) as follows:
\begin{align}
    r^\text{peak}_T &= \frac{2(T_\text{max} - T_\text{min})}{L_\text{LEO}}
\end{align}
where $T_\text{max}$ and $T_\text{min}$ are the max and min expected temperature limits in orbit, respectively, and $L_\text{LEO}$ is the typical period of a LEO orbit in minutes. The factor of two accounts for the fact that temperature increases and decreases each cycle. This rate can be estimated using SpaceX's Rideshare User Guidelines \cite{spacex_falcon} projected temperature limits and a (typical LEO orbit) period of between 90-120 minutes:
\begin{align}
    r^\text{peak}_T &= \frac{2(52^\circ C - -20^\circ C)}{\left[120 \text{min},90 \text{min}\right]}\\
    &= [1.2, 1.6] \frac{^\circ C}{\text{min}}
\end{align}
Thus, the simulated rates, whether referring to the average rate of $4.55^\circ$C/min or the peak rate of $6.74^\circ$C/min, are well in excess of expectation.

Thorough observations of the array after testing revealed no structural damage, mechanical weakness, or change to electromagnetic performance.

\subsection{Thermal Shock}
The array was subsequently subjected to repeated dunks in liquid nitrogen. This serves the purpose of verifying the antenna's survival under extreme thermal limits, repeated thermal shock, and extremely high thermal cycling rates.

Tests consisted of the following schedule:
\begin{enumerate}
    \item Insert the array into a bath of liquid nitrogen ($<-196^\circ$C)
    \item Hold for 5 seconds.
    \item Move the array from the liquid nitrogen bath to room temperature air.
    \item Hold for 25 seconds.
    \item Repeat.
\end{enumerate}
Holding periods were determined based on observation of how long it took the array to reach the desired temperature\footnote{Though liquid nitrogen and air (mostly nitrogen) have similar specific heats \cite{N2_specific_heat, air_specific_heat}, liquid nitrogen is more dense and can cool the array much faster than air can warm it.}.

This corresponds to a ``square wave," low duty-cycle stimulus at extreme temperatures representing a proportionally extreme thermal cycling rate. We performed 20 cycles over 10 minutes. 

Thorough observations of the array after testing revealed no structural damage, mechanical weakness, or change to electromagnetic performance.

\subsection{Deployment After Stowage}
Though not required for all applications, a long stowage period is typical of space missions as payloads must be prepared well in-advance of launch and may sit idle for months before (and after) launch owing to planned and/or unplanned delays in launch and/or deployment. Payloads need to survive this period without unacceptable performance deterioration. For a typical (rigid) payload, this is not a complex endeavor and involves packing the payload securely to prevent water damage, dust accumulation, etc. However, the viscoelastic characteristic of a composite can change when deformed - a phenomenon that is a function of time {\it and} temperature. Long periods of time in a stressed conformation (flattened, in this case) can warp the composite and lead to changes in array structure and, thus, electromagnetic performance. If, for example, the element does not ``pop up" to the full extent intended, the dipole arms may return to a height $<\frac{\lambda}{4}$, thus leading to destructive interference between the forward and reflected wave in the broadside direction.

\subsubsection{Time-Temperature Superposition}
\begin{figure}[t]
\centering
    \includegraphics[width=0.48\textwidth]{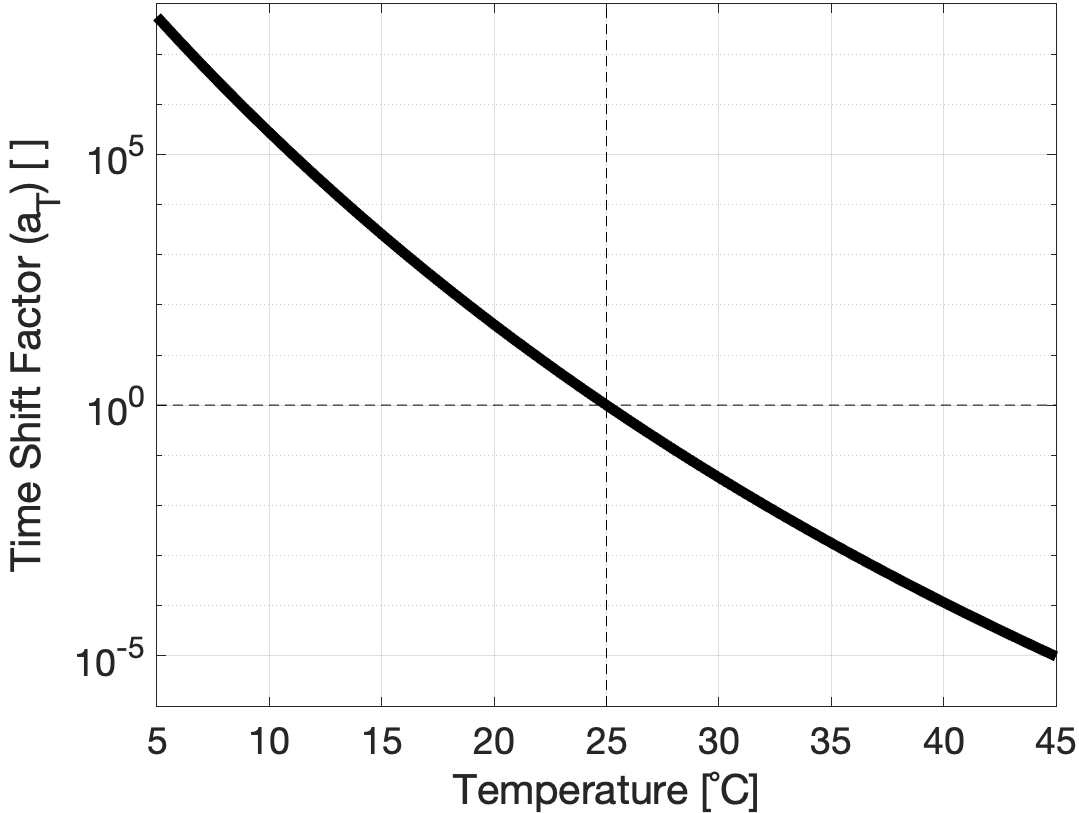}
    \caption{Time shift factor ($a_T$) for Patz-F4, where $T_0 = 25^\circ$C.}
    \label{fig:patz_time_shift}
\end{figure}

Testing deformation on the order of months is practically prohibitive. Time-temperature symmetry theory allows us to simulate the effects of long times periods by trading off larger times vs. higher temperatures; high temperatures allow us to accelerate time with regard to the composite's mechanical degradation. As discussed in \cite{patz_mastercurve}, time ``shift" relies on the assumption of the existence of a time shift factor, $a_T$, the ratio of the real time to the ``reduced time:"
\begin{align}
    a_T &= \frac{t}{t'}
\end{align}
where $T$ is the operational temperature, $t$ is the real time, and $t'$ is the ``reduced time." By definition, $a_{T_0} = 1$, which implies that time shift is relative to some reference temperature, $T_0$. To calculate $a_T$, we invoke the Williams–Landel–Ferry equation \cite{williams_landel}:
\begin{align}
    \text{log}_{10}(a_T) &= -\frac{c_1(T-T_0)}{c_2 + (T-T_0)}
\end{align}
where $c_1$ and $c_2$ are constants associated with the composite. For the relevant composite matrix (Patz-F4), $c_1$ and $c_2$ were obtained from \cite{patz_mastercurve}. The time shift factor as a function of temperature for Patz-F4 matrix is shown in Fig. \ref{fig:patz_time_shift}.

Deformation of the composite is limited by the epoxy matrix - which is structurally weaker than glass fiber. In the data below, time acceleration is computed using $c_1$ and $c_2$ associated with the {\it epoxy matrix}, as opposed to the entire composite. Because the epoxy is the limiting visco-elastic component in the composite, the tests effectively set a lower bound on the behavior of the array. For example, if the array deploys correctly after a simulated 4 months, one could conclude that the array would survive {\it at least} 4 months in stowage. Because the fibers strengthen the composite, it is possible the array would survive much longer.

It is also desirable to use physical constants associated with only the matrix, as it bypasses the requirement of generating a ``master curve" for every composite, which can be made using a large number of combinations of epoxies and fibers.

\subsubsection{Test Setup}
12 array elements (antennae) were organized into six bins of two elements each, with each bin designated for a different accelerated time period. Elements were flattened under weights (stowed), laid out on a tray, and placed in a thermal chamber set to hold at a temperature greater than room temperature. Every ten minutes, the bin of samples associated with that time (10 minutes, 20 minutes, etc.) was removed from the chamber and allowed to deploy (unflatten).

The elements' shapes were measured before and after testing to characterize changes induced by the simulated extended stowage. Shape measurement consisted of recording the deployed, vertical antenna height and deployment time\footnote{Deployment time is defined as the time between release of the antenna from a flat configuration (stowed) to the time at which it returns to a vertical position and stops vibrating.}. These metrics are most relevant to stowage-induced visco-elastic degradation, which typically presents as an inability to return to the original shape and longer deployment time(s).

The test time and temperature were chosen to simulate a year of stowage at room temperature ($25^\circ$C). At a temperature of $40^\circ$C, the epoxy matrix has a time shift factor of $1.17\times 10^{-4}$, meaning one year can be simulated in one hour. Both the short time and the moderate temperature are convenient for testing purposes. 

\begin{figure}[t]
\centering
    \includegraphics[width=0.48\textwidth]{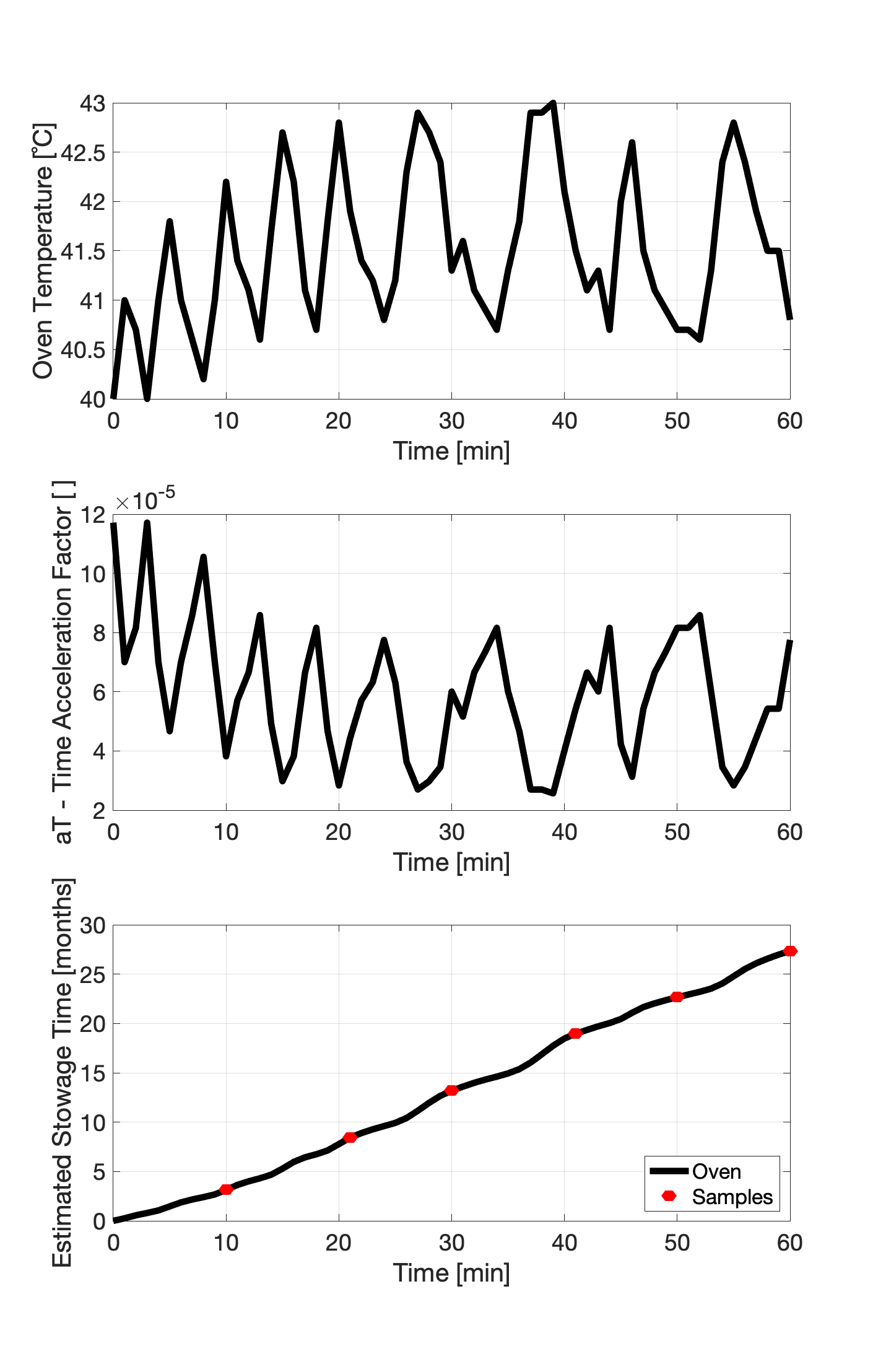}
    \caption{Top: oven temperature, $T(t)$. Middle: time shift factor, $a_T$. Bottom: cumulative reduced time, $t'(t)$.}
    \label{fig:oven_temp_shift}
\end{figure}

The thermal chamber is programmed to the desired temperature, but may not perfectly set this temperature for the duration of the experiment. Instead, it's better to estimate accelerated time using recorded temperatures from a thermal probe. These values can be processed using a generalized model for time shift, also presented in \cite{patz_mastercurve}, to better predict time acceleration:
\begin{align}
    t'(t) &= \int_{t_0}^{t}\frac{d\tau}{a_T}
\end{align}
where $t'$ is the ``reduced time" and $t$ is the real time. If $a_T$ is constant and $t_0 = 0$ (WLOG), this equation reduces to $t = a_Tt'$, as above. However, if $a_T$ is dynamic due to varying temperature, we generalize to the following:
\begin{align}
    t'(t) &= \int_{t_0}^{t} \frac{d\tau}{a_T(\tau)}\\
          &= \int_{t_0}^{t} 10^{\frac{c_1(T(\tau)-T_0)}{c_2 + (T(\tau)-T_0)}}d\tau
\end{align}
A trapezoidal Reimann sum can be used to evaluate this integral using discrete temperature readings:
\begin{align}
    t'(t) &\approx \sum_{i=1}^{N_{\tau}}\left(10^{\frac{c_1(T(\tau_i)-T_0)}{c_2 + (T(\tau_i)-T_0)}} + 10^{\frac{c_1(T(\tau_{i-1})-T_0)}{c_2 + (T(\tau_{i-1})-T_0)}}\right)\frac{\Delta\tau}{2}
\end{align}
where $N_\tau = \frac{t - t_0}{\Delta \tau}$ and $\Delta\tau$ is our sampling period (in this case: one minute). The oven temperature and corresponding cumulative reduced time are shown in Fig. \ref{fig:oven_temp_shift}.
As can be seen above, the oven temperature was consistently above the desired temperature and displayed low-amplitude oscillations owing to the bang-bang feedback mechanism of the thermal chamber, as previously discussed. Overall elevated temperature had the effect of accelerating time faster than predicted, resulting in more than double the reduced time: 27 months simulated after an hour in the oven.

\begin{figure}[t]
\centering
    \includegraphics[width=0.48\textwidth]{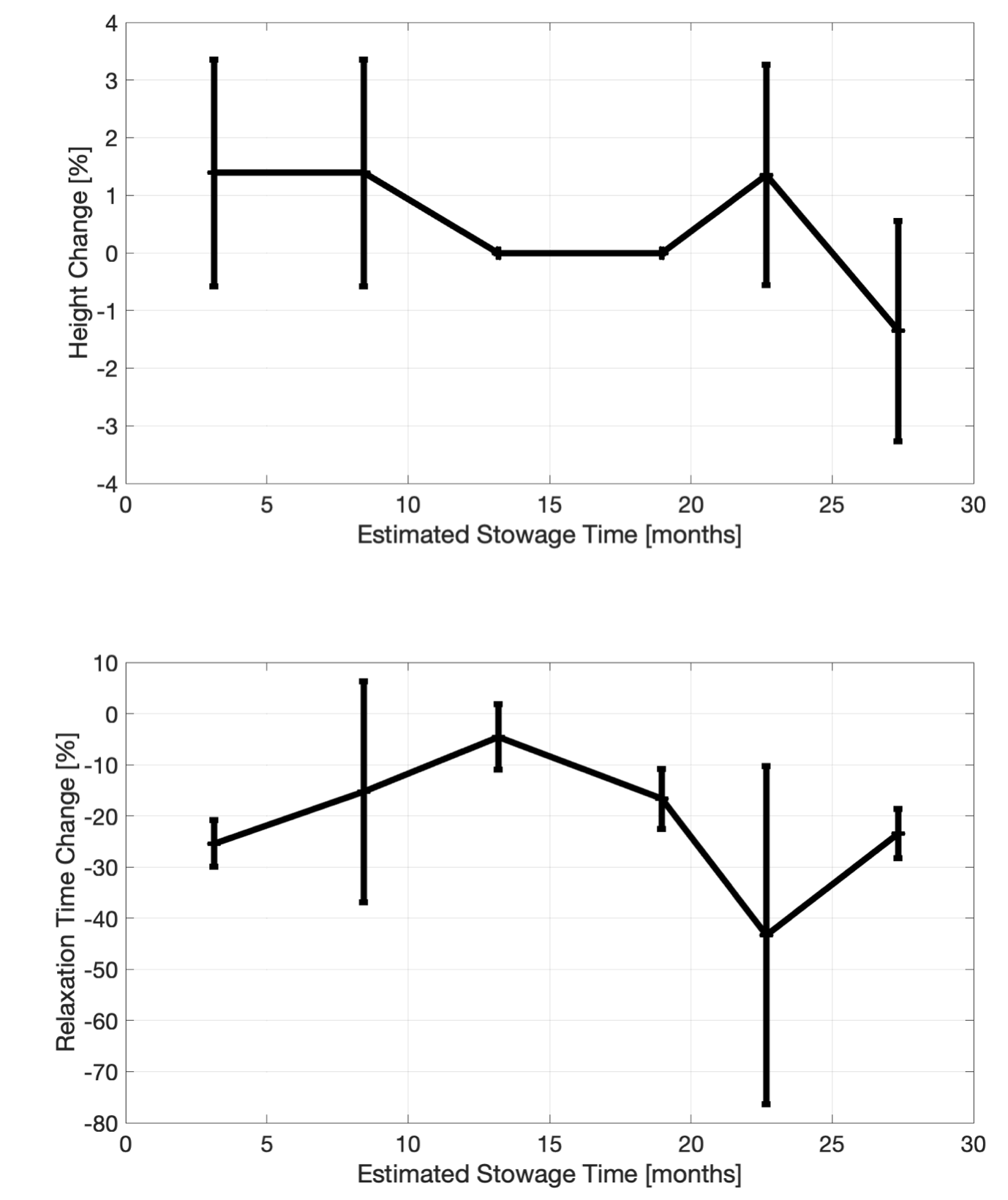}
    \caption{Changes to (upper) deployed height and deployment relaxation time (lower) as a function of (simulated) stowage time.}
    \label{fig:stowage_geometry_change}
\end{figure}

According to the two metrics - deployed height and deployment time - the antenna displayed no signs of structural damage or geometric change. As displayed in Fig. \ref{fig:stowage_geometry_change}, the deployed heights before and after the simulated stowage were identical and deployment times after simulated stowage were actually slightly faster. 

Both of these metrics indicate the array can survive {\it at least} two years in stowage with no detriment to performance. Obtaining the true upper limit on survival time would require both a longer experiment and a master curve for the entire composite and may prove unnecessary given the rarity of applications requiring a stowage period in excess of two years.

\section{Conclusion}
Presented thus far is a candidate for large, durable apertures - specifically designed for space missions but with a host of other potential applications. The arrays are light, thin, flexible, mechanically robust, and lend themselves to the large-scale automated manufacturing necessary to populate a large surface area. The arrays boast a center frequency of 10.4032 GHz with a mean bandwidth of 1.8285 GHz (17.58\%). The elements demonstrate a wide radiation pattern suitable to electronic beam steering with a half-power beam width of $110^\circ$ along both cut directions.

Moreover, the array passed the wide and demanding gamut of space-readiness testing which demonstrate the array's viability under the extreme conditions characteristic of space missions. The results from these tests also build confidence that the arrays can be deployed in other contexts with extreme conditions - in the desert or on a boat, for instance - that demand light, flexible, and durable apertures for communication and potentially wireless power transfer.

This is not the end of the road. Many design choices - the choice of epoxy, the choice of fiber, the number of layers, the curing profile - were limited and made based on availability and convenience. The co-curing method presented herein can serve as a template for a design process that allows for wide degrees of freedom in the material and processing choices. These degrees of freedom open up the possibility of tuning device performance - stiffness, mass, volume, cost - to the application's need. It is hoped that this paper will serve as the beginning of a wide-scale exploration of these techniques as methods for implementing large-scale apertures for space and other applications.

\bibliographystyle{IEEEtran}
\bibliography{main}

\end{document}